\documentclass[preprint,12pt]{elsarticle}
\usepackage[T1]{fontenc}
\usepackage{lmodern}
\usepackage{algorithm}
\usepackage{algorithmic}
\usepackage{listings}
\usepackage{multirow}

\usepackage{amssymb}

\journal{Journal of Systems Architecture}

\begin{document}

\begin{frontmatter}



\title{Accelerating Algorithms using a Dataflow Graph in a Reconfigurable System}


\author[usp1]{Jorge Luiz e Silva}
\author[usp1]{Bruno de Abreu Silva}
\author[usp1]{Joelmir Jose Lopes}
\author[UTFPR]{Antonio Carlos F. da Silva}

\address[usp1]{University of Sao Paulo, Sao Carlos, Sao Paulo, Brazil}

\address[UTFPR]{Federal Technological University of Parana, Cornelio Procopio, Brazil}

\begin{abstract}
In this paper, the acceleration of algorithms using a design of a field programmable gate array (FPGA) as a prototype of a static dataflow architecture  is discussed. The static dataflow architecture using operators interconnected by parallel buses was implemented. Accelerating algorithms using a dataflow graph in a reconfigurable system shows the potential for high computation rates. The results of benchmarks implemented using the static dataflow architecture are reported at the end of this paper. \end{abstract}


\end{frontmatter}

\section{Introduction}
\label{}

With the advent of reconfigurable computing, basically using a Field Programmable Gate Array(FPGA), researchers are trying to explore the maximum capacities of these devices, which are: flexibility, parallelism, optimization for power, security and real time applications \cite{r33,r82}.

Because of the complexity of the applications and the large possibilities to develop systems using FPGAs, many applications to convert algorithms into these devices associated with a General Purpose Processor (GPP) using high level language like C and Java is one of the challenges for researchers nowadays, especially for accelerating algorithms \cite{r78,r79}.

The main aim of this project was to accelerate the algorithms which convert parts of programs written in C language into a static dataflow model implemented in a FPGA.

This paper is organized as follows. Related work is described in section \ref{l1}. The Dataflow Graph Model is discussed in section \ref{l2}. In section \ref{l3} the Benchmarks implemented in the Dataflow graph are presented. Section \ref{l4} shows the results of the implementations. Section \ref{l5} concludes the paper and suggests future works.

\section{Related Work} \label{l1}

The dataflow graph model and its architecture was first researched in the 1970s and was discontinued in the 1990s \cite{r5,r22,r71,r72}. Nowadays, it is a topic of research once more, mainly because of the advance of technology, particulary with the advent of the FPGA \cite{r12,r71,r82}.

Because the dataflow model has an implicit parallelism and the FPGA is composed by parallel circuits, the dataflow model applied to a FPGA has the perfect combination to execute applications which also have parallelism in their execution \cite{r71}. However, as applications become more complex, software development is only possible using high level language such as C or Java \cite{r14} although only parts of the program will be executed directly into the hardware. Thus several tools have been developed to convert C into hardware using VHDL language \cite{r37,r64,r66}.

In order to analyze the data dependence, many of these systems generate an intermediate dataflow graph for pipeline instructions. The optimizations, using several techniques such as loop unrolling, are concluded and finally a reconfigurable hardware using the VHDL language is generated. The hardware generated using these tools consists of coarse grain elements or  assembler instructions for a customized processor as Picoblase or Nios from Xilinx and Altera respectively \cite{r84}.

In our approach, a fine grain instruction using VHDL to implement a static dataflow architecture, consisting of various nodes of processing elements and arcs to connect those nodes in a graph, is used to accelerate algorithms.

\section{The Dataflow Graph Model}\label{l2}

In the Asynchronous Dataflow Graph project developed by Teifel et al.   \cite{r82}, the asynchronous system is a collection of concurrent hardware processes that communicate with each other through message-passing channels. These messages consist of atomic data items called tokens. Each process can send and receive tokens to and from its environment through communication ports.  In the Teifel project, asynchronous pipelines are constructed by connecting these ports to each other using channels, where each channel is allowed only
one sender and one receiver. Since there is no clock in an asynchronous design, processes use handshake protocols to send and receive tokens via channels.

\begin{figure}[h]
\begin{center}
   \includegraphics[width=5 in]{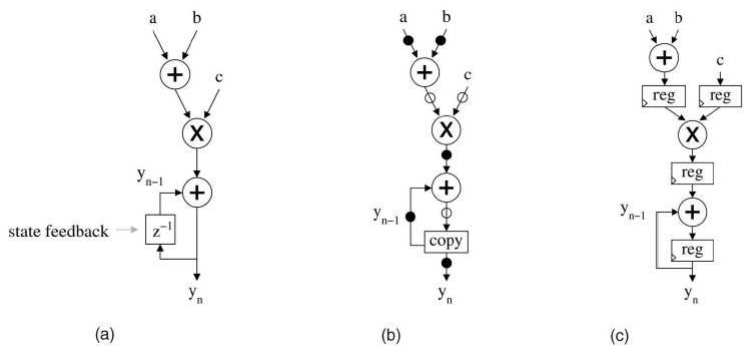}
   \caption{Computation of y{\tiny \emph{n}}=y{\tiny \emph{n-1}}+c(a+b):(a) pure dataflow graph, (b) token-based asynchronous dataflow pipeline (filled circles indicate tokens,
empty circles indicate an absence of tokens), and (c) clocked dataflow pipeline \cite{r82}.} \label{f1}
\end{center}
\end{figure}

In Fig. \ref{f1} Teifel describes an equation converted into a dataflow graph in three different situations: (a) a pure dataflow graph, (b) a token-based asynchronous dataflow pipeline and (c) a clocked dataflow pipeline.

In our project, a collection of concurrent hardware processes that communicate with each other, but using a parallel bus with bits for data and bits to control the communication in a synchronous system of communication as described in part (c) of the Fig. \ref{f1}, is also used.

\subsection{Dataflow Computations}

In the dataflow graph to accelerate algorithms project, a traditional dataflow model described in the literature, where a node is a processing element and an arc is the connection between two elements, is used \cite{r5,r12,r22,r71,r72}. A data bus and a control bus to execute the communication between the operators were implemented. The static dataflow graph model, where only one item of data can be in an arch was developed.

\begin{figure}[h]
\begin{center}
   \includegraphics[width=1.5 in]{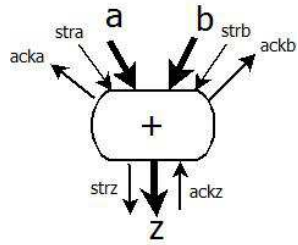}
   \caption{The basic operator with its data buses and control buses.} \label{f2}
\end{center}
\end{figure}

In Fig. \ref{f2}, a basic operator and its data buses and control buses for communication are described. The signal data  {\it a}, {\it b} and {\it z} in Fig. \ref{f2} are 16-bit data traveling through the parallel buses. The signals {\it stra}, {\it strb}, {\it strz}, {\it acka}, {\it ackb} and {\it ackz} are 1-bit control data to control communication between operators.

\begin{figure}[h]
\begin{center}
   \includegraphics[width=4 in]{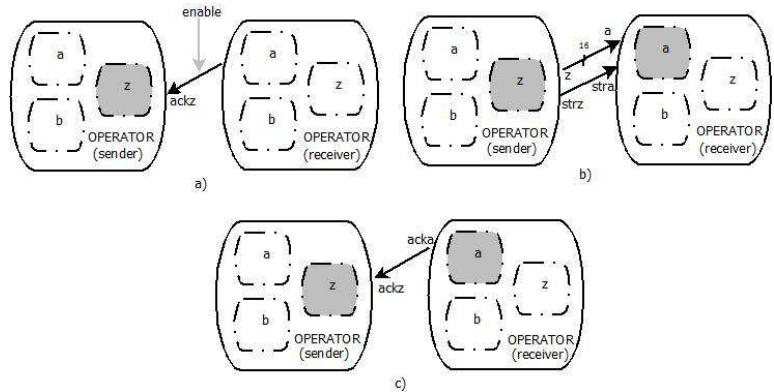}
   \caption{The communication: a) enabling the communication, b) sending an item of data, c) Acknowledging an item of data.} \label{f3}
\end{center}
\end{figure}

The communication between operators is described in Fig. \ref{f3}. As can be clearly seen in the figure, a sender operator and a receiver operator have two input data buses {\it a} and {\it b}, one output data bus {\it z} and its respective control signals  {\it stra}, {\it strb}, {\it strz}, {\it acka}, {\it ackb} and {\it ackz}. Each of the input data bus and output data bus is connected to a register to store a receiving item of data and to store a sending item of data, represented by rectangles with rounded edges {\it a, b} and {\it z} in the figure. The output data bus {\it z} from the sender operator is connected to input data bus {\it a} from the receiver operator, the output control signal {\it strz}  from the sender operator is connected to the input control signal {\it stra} from the receiver operator and the input control signal {\it ackz} from the sender operator is connected to the output control signal {\it acka} from the receiver operator.

A "logic-0" in the signal {\it ackz} informs the sender operator that the receiver operator is ready to receive data. A "logic-1' in the signal {\it ackz} informs the sender operator that the receiver operator is busy. A "logic-1" in the signal {\it stra} informs the receiver operator that an item of data is ready to be sent to it from the sender operator. A "logic-0' in the signal {\it stra} informs the receiver operator that the sender does not an item of data to be sent to it.

To initiate the communication, an {\it enable} signal with a "logic-0" to the  {\it ackz} connected to the sender, is set, Fig. \ref{f3}a. When the receiver operator is ready to receive data, a "logic-1" in the {\it stra} strobes an item of data to the input data bus {\it a} in the receiver operator, Fig. \ref{f3}b. Consequently, a "logic-0" in the {\it acka} acknowledges that the item of data {\it a} was received, Fig. \ref{f3}c.

\subsection{The Dataflow Operators}

The dataflow operators were the traditional operators described by Veen in \cite{r72}, which are: copy, non deterministic merge, deterministic merge, branch, conditional and primitive operators (add, sub, mul, div, and, or, not, etc.).

\begin{figure}[h]
\begin{center}
   \includegraphics[width=4 in]{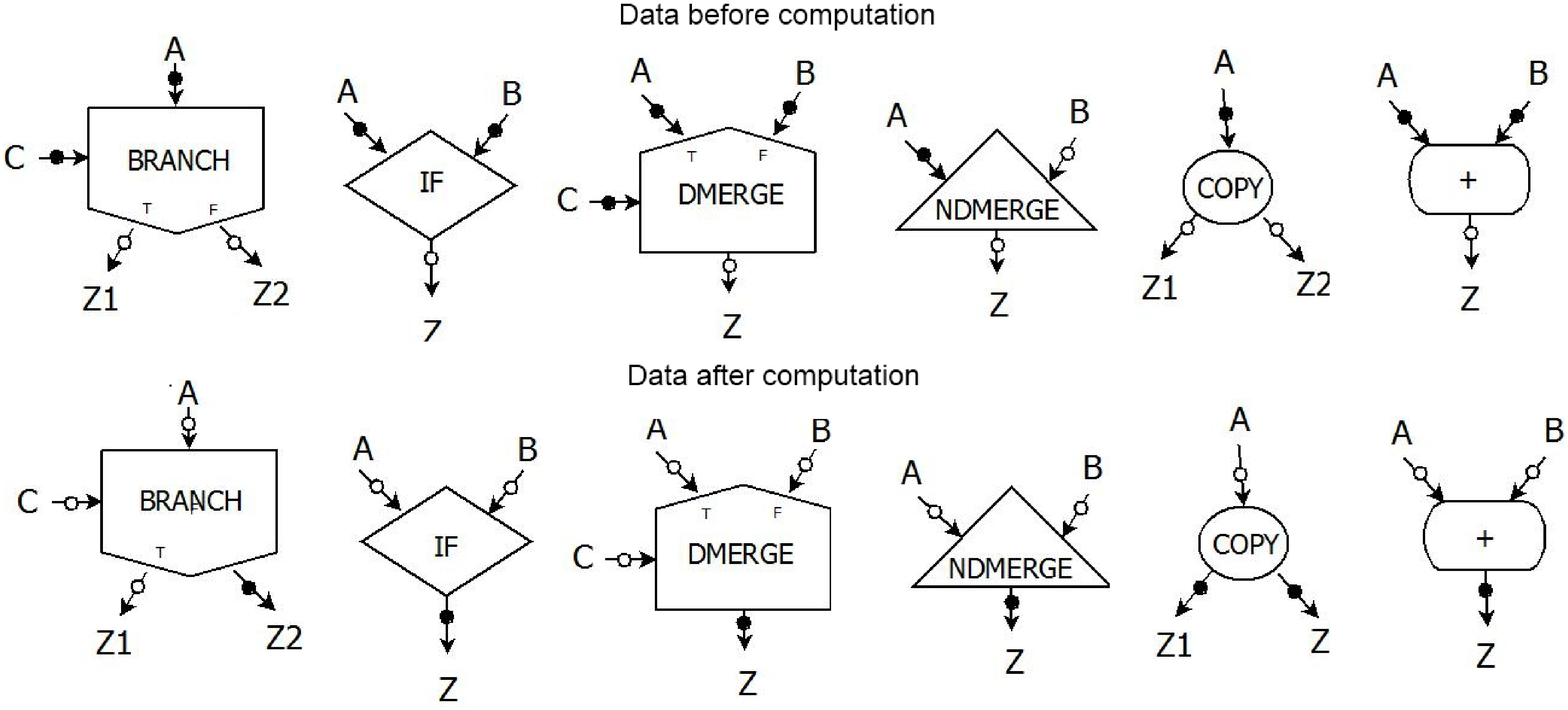}
   \caption{The operator. \cite{r82}} \label{f4}
\end{center}
\end{figure}

In order to execute the computation of an operator it is necessary that an item of data is presented in all its input buses of data. In Fig. \ref{f4}, operators are described where filled circles indicate items of data and empty circles show an absence of items of data and the situation of the operator before computation and after computation \cite{r82}.

The functional execution of dataflow operators is described below:

\begin{enumerate}
  \item {\it Copy}: This dataflow node duplicates an item of data to two receiver operators. It receives an item of data in its input data bus and copies the item of data to two output data buses.
  \item {\it Primitive}: This dataflow node receives two item of data in its input data buses, computes the primitive operation with these two items of data and generates the result sending it to the output data bus. Operators such as add, sub, multiply, divide, and, or, not, if, etc., are implemented in the same way.
  \item {\it Dmerge}: This dataflow node performs a two-way controlled data merge and allows an item of data to be conditionally read in input data buses. It receives a TRUE/FALSE item of data to decide what input data {\it a} or {\it b} respectively to send to the output data {\it z}
  \item {\it NDmerg}: This dataflow node performs a two-way not controlled data merge and allows an item of data to be read on input data buses. The first data to arrive into the Ndmerge operator from input {\it a} or {\it b} is sent to the output data {\it z}.
  \item {\it Branch}: This dataflow node performs a two-way controlled data branch and allows the item of data to be conditionally sent on to two different output buses. It receives a control TRUE/FALSE item of data to decide what output data {\it t} or {\it f} respectively to transfer the input data {\it a}.
\end{enumerate}

\subsubsection{The Basic Dataflow Operator Architecture}

\begin{figure}[h]
\begin{center}
   \includegraphics[width=4 in]{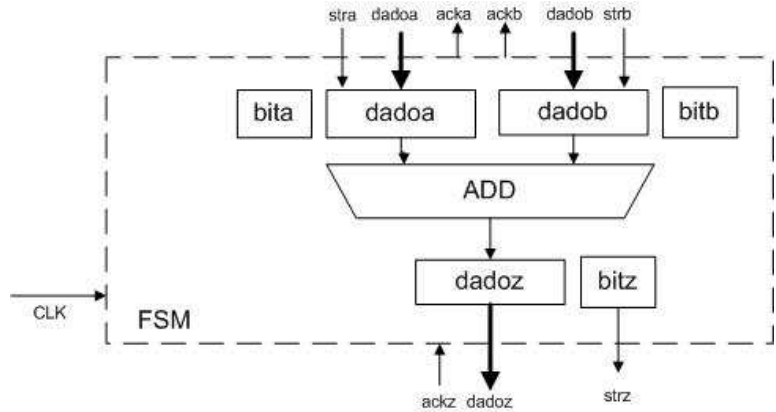}
   \caption{Datapath (RTL) Diagram of ADD Operator.} \label{f5}
\end{center}
\end{figure}

A register-transfer-level datapath (RTL) diagram for a sum (ADD) Operator is given in Fig. \ref{f5}. In the figure, the 1-bit register {\it bita} and  1-bit register {\it bitb} are  used to inform the ADD operator when the 16-bit register {\it dadoa} and/or 16-bit register {\it dadob} are filled with an item of data, respectively.

A "logic-1" in the {\it bita} or {\it bitb} informs the ADD operator that there is a item of data within {\it dadoa} or {\it dadob} respectively. A "logic-0" in {\it bita} or {\it bitb} informs the ADD operator that the {\it dadoa} or {\it dadob} is empty.

When both items of data are in the receiver operator, the ADD operator is executed and the result is filled within a 16-bit register {\it dadoz}. The 1-bit register {\it bitz} receives a "logic-1" to inform that there is a item of data to send to the next operator (the signal {\it strz} in Fig. \ref{f5}).

\begin{figure}[h]
\begin{center}
   \includegraphics[width=3.5 in]{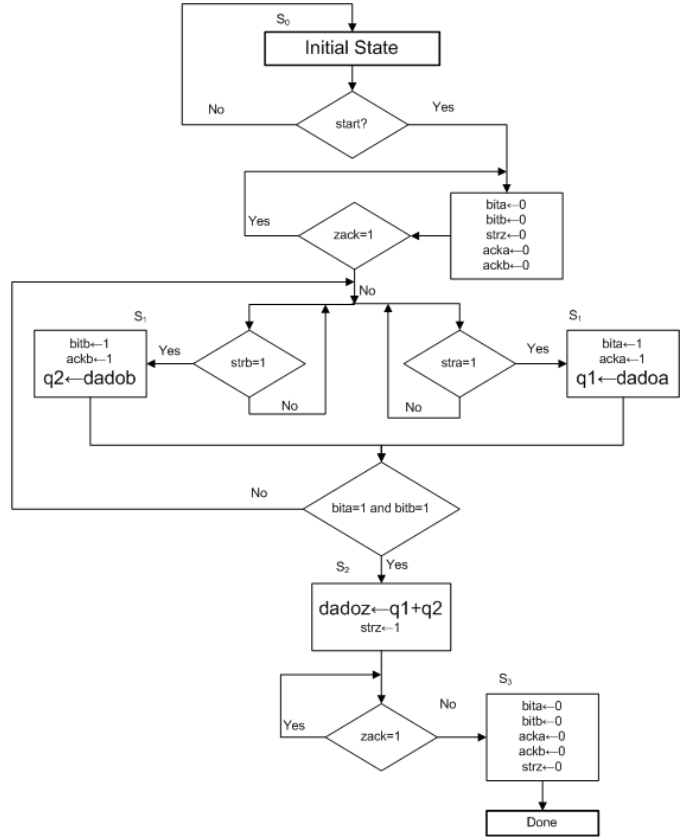}
   \caption{ASM Chart of ADD Operator.} \label{f6}
\end{center}
\end{figure}

The operation process of the ADD operator is described in the ASM chart in Fig. \ref{f6}. In the figure, there are four described states {\it S0, S1, S2 and S3}. As can be clearly seen in the figure, the initial state {\it S0} is used to initialize several signals of the operation process. In state {\it S1}, an item of data from the input data buses can be received within the operator and the correspondent bit of status can be set. Simultaneously the acknowledge signal is also set. After receiving  all the items of data, the execution of the function within the operator is started, described in state {\it S2}. Finally, in state {\it S3}, several signals of the operation process are set to "logic-0" to continue the execution process of the operator.

In the process of the operator there is a Finite State Machine (FSM) that controls each step of the execution and the communication between operators.

Although there is a clock (signal CLK in Fig. \ref{f5}), communication between operators is asynchronous because it is unpredictable when data will be sent to the next operator.

There are three different architectures of operators. One of them is already described in Fig. \ref{f5}, with two input data buses and just one output data bus. That is the case of the primitive operators {\it ADD}, {\it SUB}, {\it MUL} and {\it DIV}; the relational operators {\it IFgt}, {\it IFge}, {\it IFlt}, {\it IFle}, {\it IFeq} and {\it IFdf}; the logic operators {\it AND}, {\it OR} and {\it NOT}; and the control operator {\it NDmerge}. Another one is the control operator {\it Dmerge} with three input data buses and just one output data bus. Finally the last one, the control operator {\it Branch} with two input data buses and two output data buses.

\section{The Benchmarks Implemented in the Dataflow Model}\label{l3}

The benchmarks implemented in the dataflow model were: Fibonacci, Max, Dot prod, Vector sum, Bubble sort, and Pop count \cite{r83}.

\begin{figure}[h]
\begin{center}
   \includegraphics[width=5 in]{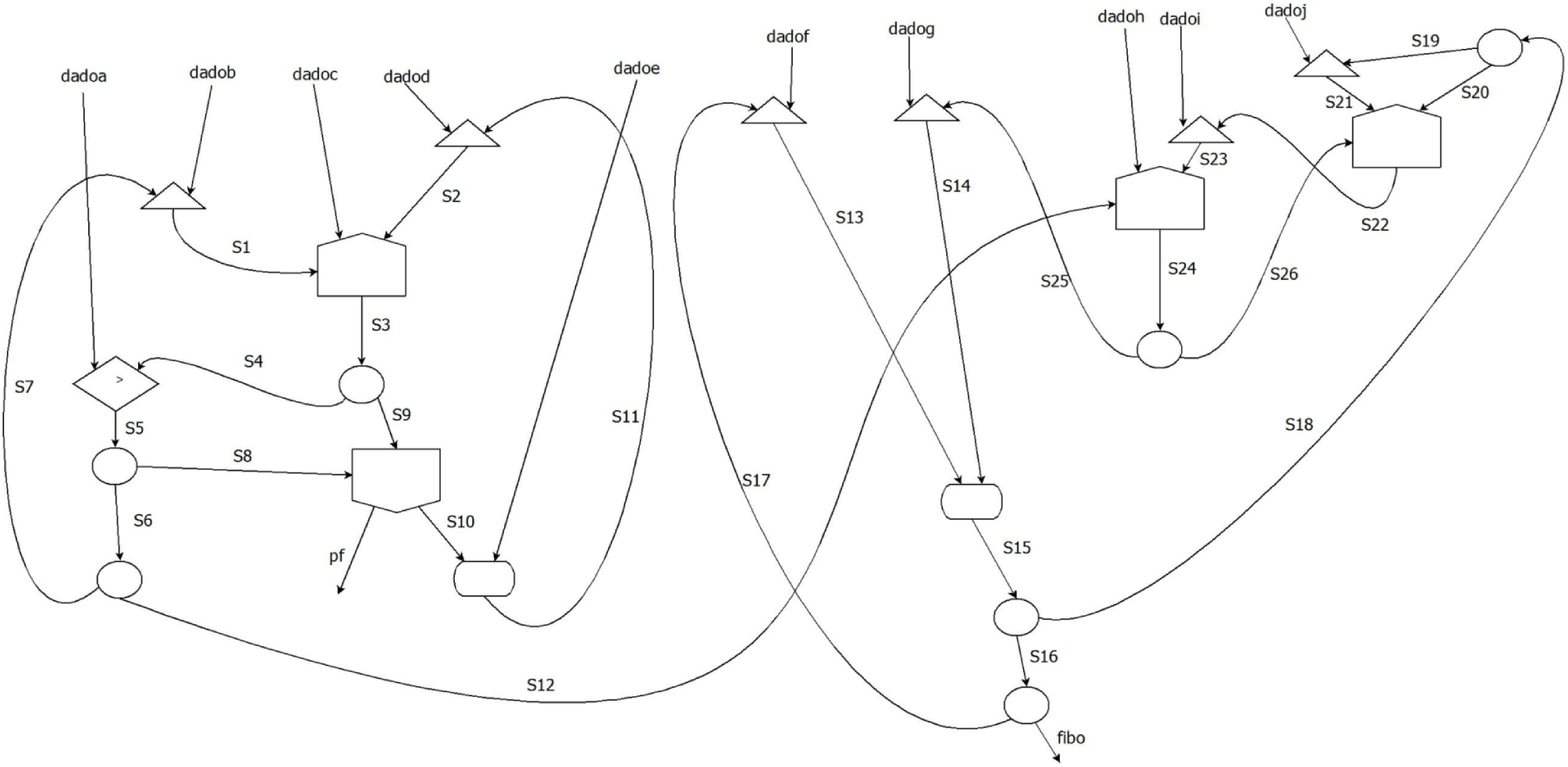}
   \caption{The Fibonacci algorithm described in Dataflow Graphics.} \label{f7}
\end{center}
\end{figure}

To convert the benchmark algorithms into a VHDL, each benchmark was described as a dataflow graph, them an assembler language was used to convert the dataflow graph into a VHDL. The Fibonacci algorithm was described just to illustrate the process to convert an algorithm into a VHDL. The others algorithms were processed in the same way. The Fibonacci algorithm is described in Algorithm \ref{alg1} and its dataflow graph is described in Fig. \ref{f7}.

\begin{algorithm}
\caption{Calculate Fibonacci}
\label{alg1}
\begin{algorithmic}
\STATE $first \Leftarrow 0$
\STATE $second \Leftarrow 1$
\STATE $tmp \Leftarrow 0$
\FOR{$i=0$ to $n$}
\STATE $tmp \Leftarrow first + second$
\STATE $first \Leftarrow second$
\STATE $second \Leftarrow tmp$
\ENDFOR
\end{algorithmic}
\end{algorithm}

As can be clearly seen in Fig. \ref{f7}, there are two parts in the dataflow graph: one of them is located on the left side of the figure and controls the loop with index {\it i}; on the right side of the figure the implementation of the Fibonacci sequence is described.

As the dataflow graph consist of nodes and arcs, each node represents an operator and each arc represents the communications between two operators. In Fig. \ref{f7}, a label is attributed to each arc in the dataflow graph. As arcs represent the communication between two operators, the parallel data bus for items of data and the control data bus for control the communications are included in the label representations. The assembler language uses the name of the operator and its label arcs to convert the dataflow graph into a VHDL. The assembly language for Fibonacci dataflow graph is described in Listening \ref{7000}.

As can be clearly seen in Listening \ref{7000}, several node operators and their input and output arcs are listed. Labels used to connect nodes operators are described initializing with the {\it s} character followed by a number and the others are input or output data signals. The labels
{\it dadoa}, {\it dadob}, {\it dadoc}, {\it dadod}, {\it dadoe}, {\it dadof}, {\it dadog}, {\it dadoh}, {\it dadoi} and {\it dadoj} are input data signals used to initialize data for the Fibonacci dataflow graph and the labels {\it pf} and {\it fibo} are output data signals to inform the result of the Fibonacci sequence. Specifically for the Fibonacci sequence, {\it dadoa} receives the {\it n} Fibonacci argument and {\it fibo} is the result of the {\it n} Fibonacci argument.

{\tiny
\lstset{language=VHDL}
\lstset{commentstyle=\textit}
\lstset{backgroundcolor=,rulecolor=}
\begin{lstlisting}[frame=tb,caption= {\it The Assembler Language for Fibonacci Dataflow Graph},label=7000]{somecode}
1.ndmerge s7,dadob,s1;
2.dmerge s2,dadoc,s1,s3;
3.ndmerge dadod,s11,s2;
4.gtdecider dadoa,s4,s5;
5.copy s3,s4,s9;
6.copy s5,s6,s8;
7.branch s9,s8,s10,pf;
8.copy s6,s7,s12;
9.add s10,dadoe,s11;
10.ndmerge s17,dadof,s13;
11.ndmerge dadog,s25,s14;
12.ndmerge dadoi,s22,s23;
13.ndmerge dadoj,s19,s21;
14.copy s18,s19,s20;
15.dmerge s23,dadoh,s12,s24;
16.dmerge s20,s21,s26,s22;
17.copy s24,s25,s26;
18.add s13,s14,s15;
19.copy s15,s16,s18;
20.copy s16,s17,fibo;
\end{lstlisting}
}

%


\section{Experimental Results}\label{l4}

The benchmarks were implemented using a (7v285tffg1157-3) Virtex FPGA from Xilinx and synthesized in ISE 13.1 and the results were compared with the same benchmarks implemented in C-to-Verilog and LALP described in \cite{r83} that were implemented using a (EP1S10F780C6) Stratix FPGA from Altera and synthesized in Quartus II 6.1.

\begin{table*}[h]
\begin{center}
\tiny
\caption{The results of implementation for Benchmarks \label{dados1}}
\begin{tabular}{|l|l|l|l|l|l|} \hline
\multirow{8}{*}{C-to-Verilog} & & & & & \\
& Benchmarks & FF & LUT & Slices & Mas Freq.\\ \hline
& Buble Sort & 2353 & 2471 & 971 & 239.45\\
& Dot prod & 758 & 578 &285 & 249.36\\
& Fibonacci & 73 & 108 &69 & 297.81\\
& Max vector & 496 & 392 & 164 & 435.9\\
& Pop count & 1023 & 872 & 384 & 411.22\\
& Vector sum & 177 & 113 & 34 & 546.538\\ \hline
\multirow{7}{*}{LALP} &&&&& \\
&Buble Sort&219&105&79&353.16\\
&Dot prod&97&69&32&213.14\\
&Fibonacci&104&41&30&505.08\\
&Max vector&50&39&20&484.97\\
&Pop count&350&215&115&503.73\\
&Vector sum&------&------&------&------\\ \hline
\multirow{7}{*}{Algorithm Accelerator} &&&&&\\
&Buble Sort&85&485&712&613.685\\
&Dot prod&323&362&542&613.685\\
&Fibonacci&72&482&755&612.108\\
&Max vector&80&425&598&613.685\\
&Pop count&79&453&684&613.685\\
&Vector sum&52&284&419&613.685\\ \hline
\end{tabular}
\end{center}
\end{table*}

\begin{figure}[h]
\begin{center}
   \includegraphics[width=5.5 in]{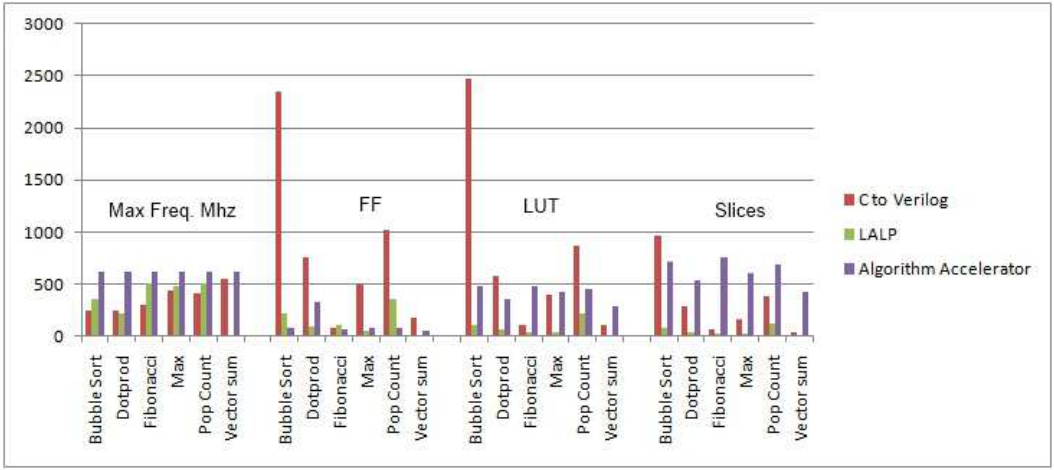}
   \caption{comparing the Benchmarks} \label{f8}
\end{center}
\end{figure}

In Table \ref{dados1}  the results of implementations for each benchmark in C-to-Verilog, LALP and Acceleration Algorithms are described. In Fig. \ref{f8}, a synthesis of the results is described.

As can be clearly seen in Fig. \ref{f8}, the Acceleration Algorithms occupy less Flip Flops (FF) than the C-to-Verilog system, but more than the LALP system, for all the benchmarks. For LUT occupancy, the Acceleration Algorithms occupy less LUTs than the C-to-Verilog system, except for the Fibonacci, Max and Vector sum benchmarks, but more than the LALP system, also for all the benchmarks. In the Slices occupancy, the Acceleration Algorithms occupy more slices than the C-to-Verligo and the LALP system (except for the Bubble sort benchmark). Finally, for Maximum Frequency, the Acceleration Algorithms had more speed than the other two systems.


\section{Conclusion and Future Work}\label{l5}

Accelerating Algorithms, by and large, occupy more space within the FPGA than the C-to-Verilog and the LALP system.  However, accelerating algorithms have more speed than the other two systems, although the main aim in this project was to validate the implementation model likely to convert algorithms into the dataflow graph and into a VHDL. Taking this into account, accelerating algorithms become one more solution for parallelism in FPGA. The benchmarks used in this paper basically perform operations using vectors, but it is very important to explore the maximum parallelism of the dataflow graph using real parallel applications. Future work would be to develop a module to convert C directly into a VHDL, associated with the FPGA and  to implement a dynamic dataflow model to obtain a better performance than the static model implemented in this paper.

\section*{References}





\bibliographystyle{elsarticle-harv}
\bibliography{elsevier_jor}







\end{document}